\documentclass[fleqn,12pt,twoside]{article}
\usepackage{espcrc1}

\usepackage{graphicx}
\usepackage[figuresright]{rotating}

\newcommand{\AmS}{{\protect\the\textfont2
  A\kern-.1667em\lower.5ex\hbox{M}\kern-.125emS}}

\title{Two-pion exchange NN potential from Lorentz-invariant 
$\chi$EFT\thanks{
Notice: Authored by Jefferson Science Associates, LLC under U.S. DOE 
Contract No. DE-AC05-06OR23177. The U.S. Government retains a 
non-exclusive, paid-up, irrevocable, world-wide license to publish or 
reproduce this manuscript for U.S. Government purposes.}
}

\author{\underline{R. Higa}\address{Jefferson Laboratory, 12000 Jefferson 
Avenue MS12H2, Newport News, VA 23606, USA}%
        \thanks{Present address: Helmholtz-Institut f\"ur Strahlen- 
und Kernphysik, Universit\"at Bonn, Nu\ss allee 14-16, Bonn, Germany, 
53115. {\tt Email:} higa@itkp.uni-bonn.de.},
        M.R. Robilotta\address{Physics Institute, University of 
S\~ao Paulo, C.P. 66318, 05315-970, S\~ao Paulo, SP, Brazil},
        C.A. da Rocha\address{N\'ucleo de Pesquisa em Bioengenharia, 
Universidade S\~ao Judas Tadeu, Rua Taquari, 546, 03166-000, 
S\~ao Paulo, SP, Brazil}}
       
\begin{document}

% typeset front matter
\maketitle

\begin{abstract}
We outline the progress made in the past five years by the S\~ao Paulo 
group in the development of a two-pion exchange nucleon-nucleon 
potential within a Lorentz-invariant framework of (baryon) chiral 
perturbation theory. 
\end{abstract}

%%%%%%%%
\section{Motivation: the problem in the heavy baryon formalism}

\begin{figure}[htb]
\begin{minipage}[t]{75mm}
\begin{center}
\includegraphics[width=45mm]{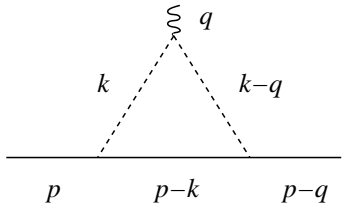}
\end{center}
\end{minipage}
\hspace{\fill}
\begin{minipage}[t]{80mm}
\vspace{-36mm}
\caption{The triangle diagram that contributes to the nucleon scalar 
form factor. The solid, dashed, and wiggly lines represent, respectively, 
the nucleon, the pions, and an external scalar-isoscalar source.}
\label{fig:fig1}
\end{minipage}
\end{figure}

In the works of Becher and Leutwyler \cite{BL} it was shown that the 
convergence of the chiral expansion of the nucleon scalar form factor, 
driven by the triangle diagram of Fig.~\ref{fig:fig1}, is a delicate 
issue for values of the momentum transfer $t$ near $(2m_{\pi})^2$ due 
to the presence of an anomalous threshold, {\em i.e.} a branch point in 
the second Riemann sheet right below the two-pion threshold. Such a 
singularity is completely neglected in heavy baryon chiral perturbation 
theory (HB$\chi$PT), and can only be recovered by resumming the heavy 
baryon series to all orders. In order to understand this problem one can 
start from the spectral representation of the triangle integral, 

\begin{equation}
\gamma(t)=\frac{1}{\pi}\int_{4m_{\pi}^2}^{\infty}\frac{dt'}{(t'\!-\!t)}\,
\mbox{Im}\gamma(t')\,,
\quad\mbox{where}\quad
\mbox{Im}\gamma(t')\simeq \frac{\theta(t'\!-\!4m_{\pi}^2)}{16\pi m_N\sqrt{t'}}\,
\arctan\frac{2m_N\sqrt{t'\!-\!4m_{\pi}^2}}{t'\!-\!2m_{\pi}^2}\,.
\end{equation}

%\vspace{-3mm}

Formally, the argument $x=2m_{N}\sqrt{t'-4m_{\pi}^2}/(t'-2m_{\pi}^2)$
is counted as order $q^{-1}$ and the HB expansion yields 
$\tan^{-1}x= \pi/2 - 1/x + 1/3x^3+\cdots $, but this is valid only in the
domain $|x|\geq 1$. The first two terms reproduce the HB result for
the triangle graph,

\begin{eqnarray}
\gamma(-q^2)|_{HB}&=&\frac{1}{16\pi^2m_{N}m_{\pi}}\int_{4m_{\pi}^2}^{\infty}
\frac{dt'}{(t'+q^2)}\,\frac{1}{\sqrt{t'}}
\left[\frac{\pi}{2}-\frac{(t'-2m_{\pi}^2)}{2 m_{N}
\sqrt{t'\!-\!4m_{\pi}^2}}\right]
\nonumber\\[2mm]
&=&\frac{1}{16\pi^2m_{N}m_{\pi}}\left[2\pi m_{\pi}\,A(q)+
\frac{m_{\pi}}{m_{N}}\,
\frac{(2m_{\pi}^2+q^2)}{(4m_{\pi}^2+q^2)}\,L(q)\right]\,,
\label{eq:new03}
\end{eqnarray}

\noindent where $q=|{\mbox{\boldmath $q$}}|$, and $L(q)$ and $A(q)$ are
the usual HB loop functions,

\begin{equation}
L(q)=\frac{\sqrt{4m_{\pi}^2+q^2}}{q}\,\ln
\frac{\sqrt{4m_{\pi}^2+q^2}+q}{2m_{\pi}}\,,
\qquad\qquad
A(q)=\frac{1}{2q}\,\arctan\frac{q}{2m_{\pi}}\,.
\end{equation}

However, it does not take into consideration the case $|x|<1$, where
$t'$ gets closer to $4m_{\pi}^2$. This region determines the long distance
behavior of the triangle diagram, as can be seen by its representation
in configuration space~\cite{HR03},

\begin{equation}
\Gamma(r)= \frac{1}{\pi}\int_{4m_{\pi}^2}^{\infty}dt'\int
\frac{d^3q}{(2\pi)^3}\;e^{-i{\mbox{\boldmath $q$}}\cdot
{\mbox{\boldmath $r$}}}\;\frac{\mbox{Im}\gamma(t')}{t'+q^2}=
\frac{1}{4\pi^2}\int_{4m_{\pi}^2}^{\infty}dt'\;\frac{e^{-r\sqrt{t'}}}{r}\;
\,\mbox{Im}\gamma(t')\,.
\end{equation}

Clearly one sees that, in order to have a good asymptotic
description of $\Gamma(r)$, one needs a decent representation for
$\mbox{Im}\gamma(t')$ near $t'=4m_{\pi}^2$, which cannot be provided 
by HB$\chi$PT.

We want to stress that the expansion in $1/m_N$ of our two-pion exchange 
nucleon-nucleon potential (TPEP) should, in principle, recover the 
expressions from HB$\chi$PT, and we used this fact as a cross-checking 
of our calculation (see next section). However one must keep in mind that, 
due to the problem described above, such an expansion should not be done. 

%%%%%%%%
\section{Comparison of HB- and RB-$\chi$PT results}

The technical details in the evaluation of our TPEP is described 
in Refs.~\cite{HR03,HRR04,RH04}. Our loop integrals are 
calculated relativistically using dimensional regularization in the 
$\overline{\rm MS}$ scheme, and we showed that they become almost 
identical as the infrared-regularized results in configuration space 
for distances above 1fm~\cite{HRR04}. Because we are constructing the 
kernel of the interaction in a relativistic way one has to deal with 
the subtraction of the iterated one-pion exchange, with the 
intermediate two-nucleon propagating only with positive energy. The 
subtraction of this contribution is not unique, and in \cite{HR03,HRR04} 
we adopted the Blankenbecler-Sugar prescription. 

The $1/m_{N}$ expansion of our TPEP and comparison with the HB results 
were shown in \cite{HR03}, where we initially found 14 different terms out 
of lengthy expressions, which will not be reproduced here. We revised our 
calculations, in particular the two loop contributions~\cite{RH04,HRR06}, 
and with the corrected expressions these differences dropped down to 9 
terms. Their origins are now better understood: six of them come from the 
prescription for the subtraction of the iterated one-pion exchange, and 
using the same procedure of Ref.~\cite{KBW97} (see~\cite{RH04}) those 
differences went away. The remaining three come from two loop diagrams, 
and the reason for such a discrepancy is harder to access. It is quite 
possible that they come from the way the one-loop counterterms in the 
$\pi$N amplitude were renormalized \cite{K01-2l,HRR06}. Numerically 
they are not significant~\cite{RH04}, but if one aims at increasing 
precision this technical issue may have to be revisited. 

From now on we will ignore these differences and focus only on the effect 
of the $1/m_{N}$ expansion. To regularize the short-distance divergence 
one considers a phenomenological cutoff of the Argonne V14-V18 type, 

\begin{equation}
\big[1-\exp(-cr^2)\big]^4\,,
\label{eq:cutoff}
\end{equation}

\noindent with $c=2\mbox{fm}^{-2}$, and in the remaining of this section 
one adopts the $\pi$N LECs from Entem and Machleidt~\cite{EM02} 
(to be discussed in the next section). The NN potential is 
further supplemented with the usual charge-dependent one-pion exchange. 

\begin{figure}[htb]
\begin{minipage}[t]{77mm}
\begin{center}
\includegraphics[width=55mm]{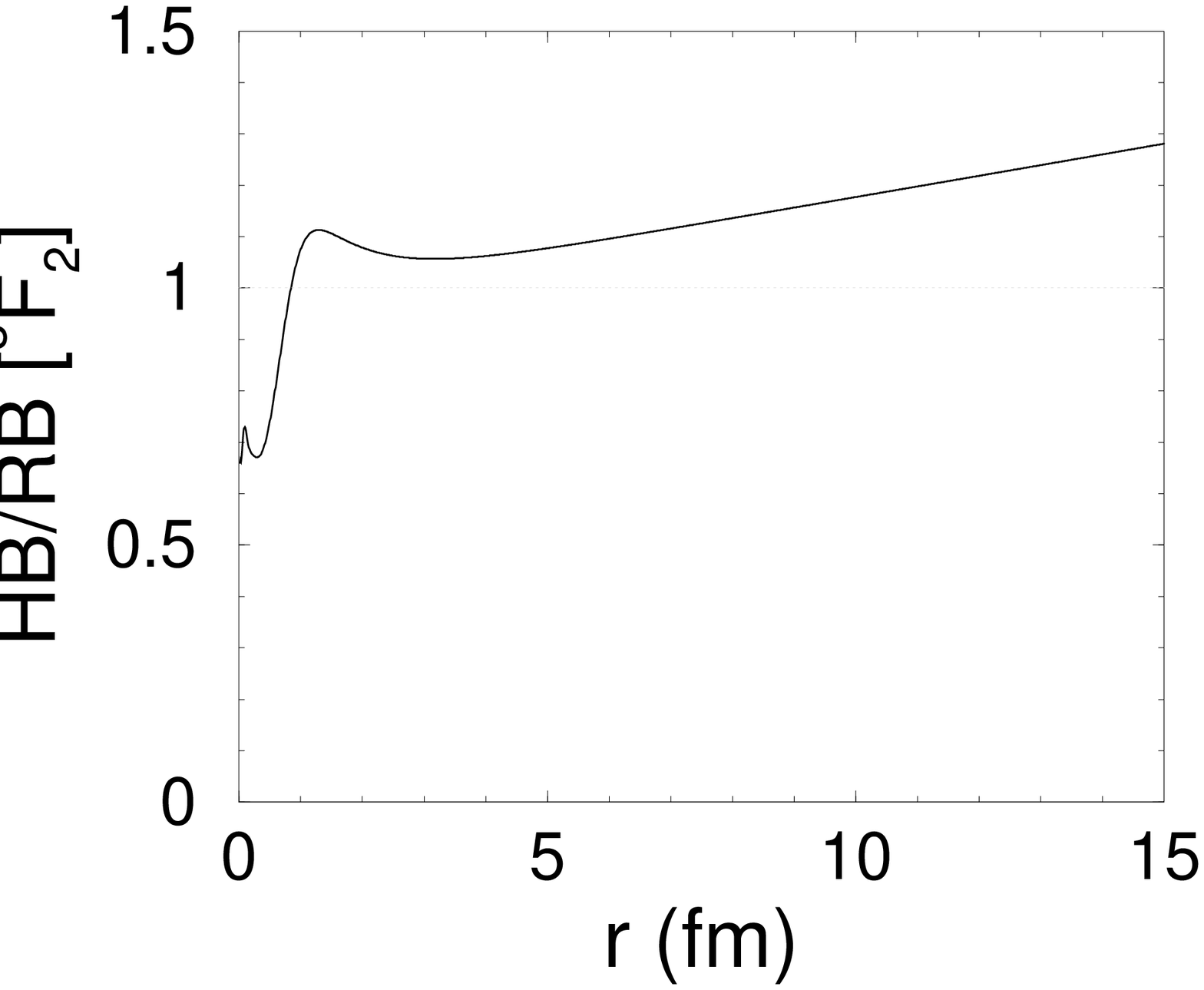}
\end{center}
\vspace{-13mm}
\caption{${}^3F_2$ partial wave projection of the ratio HB- over the 
RB-$\chi$PT TPEP.}
\label{fig:fig2}
\end{minipage}
\hspace{\fill}
\begin{minipage}[t]{77mm}
\begin{center}
\includegraphics[width=55mm]{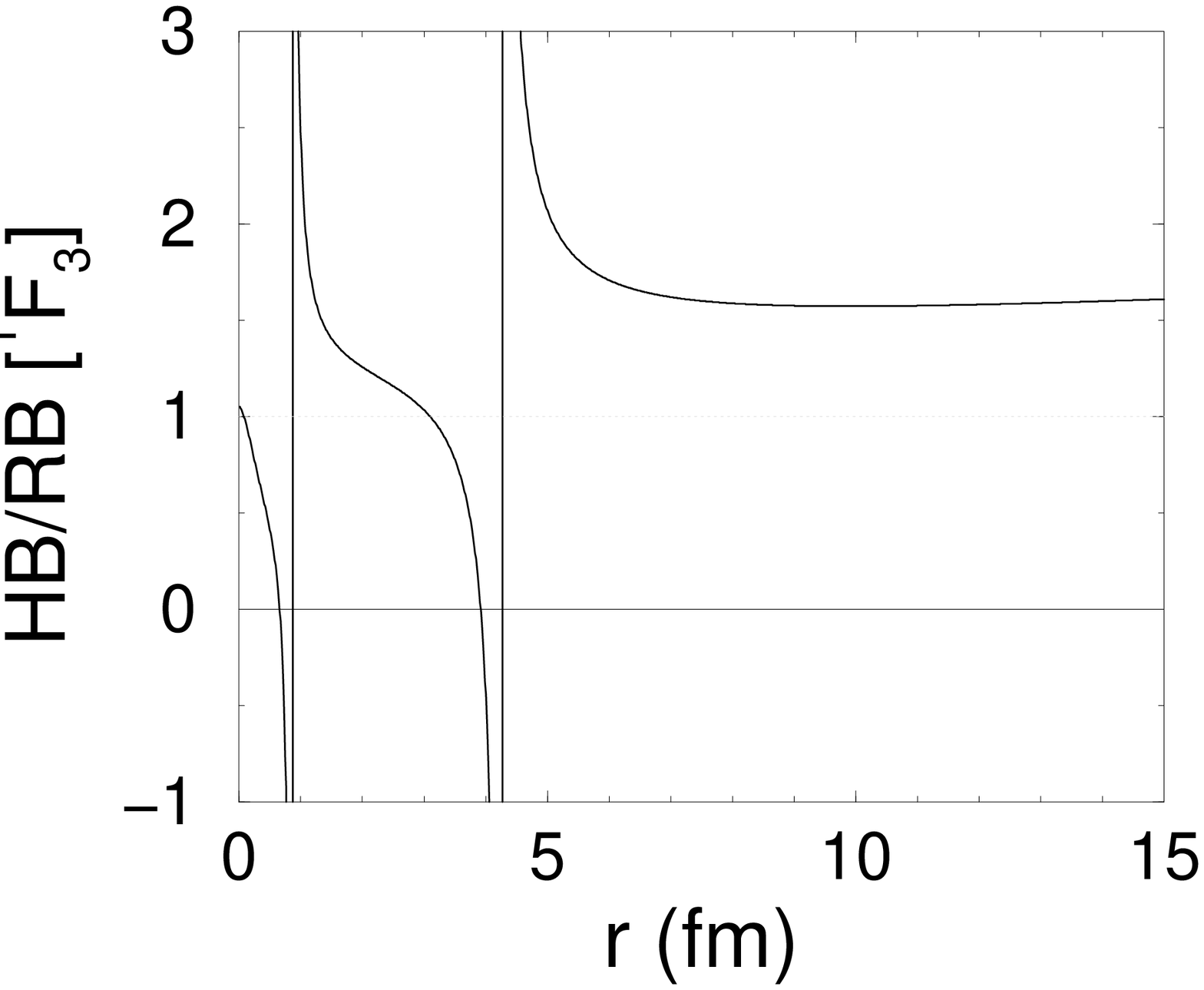}
\end{center}
\vspace{-13mm}
\caption{${}^1F_3$ partial wave projection of the ratio HB- over the 
RB-$\chi$PT TPEP.}
\label{fig:fig3}
\end{minipage}
\end{figure}

Figs.~\ref{fig:fig2} and \ref{fig:fig3} show the ratios of the $1/m_N$ 
expanded (HB) over the non-expanded (RB) TPE potentials for ${}^3F_2$ and 
${}^1F_3$ partial waves and they represent, respectively, what one observes 
for the total isospin $T=1$ and $T=0$ channels. For $T=1$ waves 
the ratio follows the behavior of the loop integrals \cite{HRR04}, with 
differences of $\sim$ 20\% at $r=10$fm and increasing with the distance. 
On the other hand, the $T=0$ case shows a sizeable factor of 1.7 already 
at $r\sim 6$fm. A numerical investigation suggests that a significant 
cancellation happens between the isoscalar and isovector components in this 
region~\cite{RH04}, therefore amplifying the difference between the HB and 
RB results. 

%%%%%%%%
\section{LECs and phase shifts}

Table \ref{tab:tab1} shows the values of the $\pi$N LECs considered in 
Ref.~\cite{RH04}, and there you can also find a short description on the 
extraction of each related work. The important observation to be made is 
the large spread in the central value of these constants, usually extracted 
from $\pi$N scattering with large uncertainties, and the question to ask is 
whether peripheral NN scattering is able to restrict some of these sets 
of LECs. 

\begin{table}[!htb]
\caption{Values for the $\pi$N LECs from the $O(q^2)$ Lagrangian, used 
in Ref.~\cite{RH04}.}
\label{tab:tab1}
\begin{center}
\begin{tabular} {|c|c|c|c|c|c|}
\hline
LEC & EM \cite{EM02} & Moj\v zi\v s \cite{Moj98} & BM \cite{BM00} & 
FMS (fit 1) \cite{FMS98} & Nijmegen \cite{Nij03} \\ \hline
$c_1$ & -0.81 & -0.94 & -0.81 & -1.23 & -0.76 \\ \hline
$c_2$ &  3.28 &  3.20 &  8.43 &  3.28 &  3.20 \\ \hline
$c_3$ & -3.40 & -5.40 & -4.70 & -5.94 & -4.78 \\ \hline
$c_4$ &  3.40 &  3.47 &  3.40 &  3.47 &  3.96 \\ \hline
%\hline
%$\bar d_1+\bar d_2$       &  3.06 &  2.40 &  3.06 &  3.06 &  2.40 \\ \hline
%$\bar d_3$                & -3.27 & -2.80 & -3.27 & -3.27 & -2.80 \\ \hline
%$\bar d_5$                &  0.45 &  1.40 &  0.45 &  0.45 &  1.40 \\ \hline
%$\bar d_{14}-\bar d_{15}$ & -5.65 & -6.10 & -5.65 & -5.65 & -6.10 \\ \hline
\end{tabular}
\end{center}
\end{table}

\vspace{-3mm}
\begin{figure}[!htb]
\begin{minipage}[t]{75mm}
\begin{center}
\includegraphics[width=65mm]{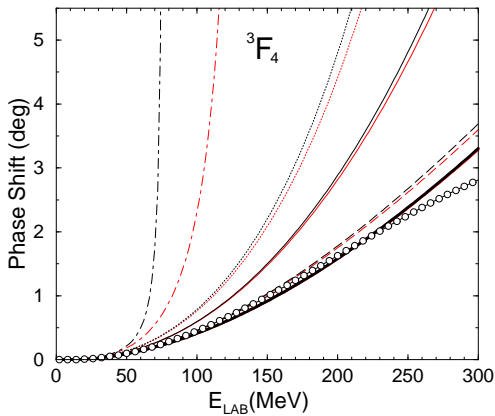}
\end{center}
\end{minipage}
\hspace{\fill}
\begin{minipage}[t]{80mm}
\vspace{-63mm}
\caption{${}^3F_4$ phase shift predictions using the LECs in table 
\ref{tab:tab1}. Solid, thick line: Entem and Machleidt~\cite{EM02}, 
dotted: Moj\v zi\v s~\cite{Moj98}, dashed: B\"uttiker and 
Mei\ss ner~\cite{BM00}, dot-dashed: Fettes {\em et.al.}~\cite{FMS98}, 
solid, thinner line: Nijmegen~\cite{Nij03}. The light and dark curves 
correspond to the HB and RB expressions, respectively, and the circled 
line, to the Nijmegen partial wave analysis~\cite{nnonline}.}
\label{fig:fig4}
\end{minipage}
\end{figure}
%\vspace{-3mm}

In Fig.~\ref{fig:fig4} we show the ${}^3F_4$ phase shifts for the LECs in 
table~\ref{tab:tab1}. Despite from what one sees in Figs.~\ref{fig:fig2} 
and \ref{fig:fig3}, the difference between HB and RB predictions for 
phase shifts is quite small, except for values of LECs from Fettes 
{\em et.al.} and Moj\v zi\v s, which are not consistent with Nijmegen's 
partial wave analysis~\cite{nnonline} (PWA) anyway. The reason is that 
the potentials differ significantly after $r\sim 5$fm, where the TPEP 
is already too small and gets washed away by the OPEP. 

Concerning the LECs, our results show more sensitivity to the 
constant $c_3$, which controls the strength of the attractive, central 
scalar-isoscalar potential. It can be inferred from the values of Fettes 
{\em et.al.} and Moj\v zi\v s (and a bit less from Nijmegen), which 
have larger values for $|c_3|$ and produces larger, positive contributions 
to the phase shifts. We also varied the cutoff parameter $c$ in 
Eq.~\ref{eq:cutoff} between 1.5 and 2.5 fm${}^{-2}$ and observed very 
small variations and good overall agreement with PWA using the LECs from 
Entem and Machleidt, and slightly bigger variations and disagreement in 
some waves for values from B\"uttiker and Mei\ss ner. In the case of 
Nijmegen, agreement with PWA is possible in some waves only with a cutoff 
as low as 1.0fm${}^{-2}$, and keeping this as a lower limit for $c$, 
no agreement with PWA is reached for LECs from Fettes {\em et.al.} or from 
Moj\v zi\v s. This indicates that peripheral nucleon-nucleon scattering 
favors smaller absolute values for the LEC $c_3$. 

\begin{figure}[!htb]
\begin{minipage}[t]{75mm}
\begin{center}
\includegraphics[width=65mm]{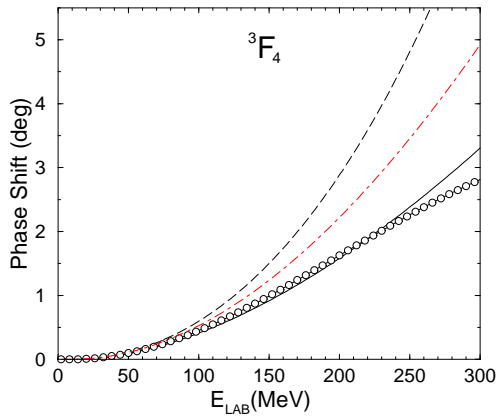}
\end{center}
\end{minipage}
\hspace{\fill}
\begin{minipage}[t]{80mm}
\vspace{-58mm}
\caption{${}^3F_4$ phase shift results for the $O(q^4)$ chiral potential 
with LECs from Entem and Machleidt~\cite{EM02} (solid curve), $O(q^3)$ 
(dashed, light curve) and $O(q^4)$ (dashed, dark curve) chiral potential 
with LECs from the Nijmegen group~\cite{Nij03}, and 
comparison with Nijmegen partial wave analysis~\cite{nnonline} (circles). }
\label{fig:fig5}
\end{minipage}
\end{figure}

A question that arises is why the LECs extracted by the Nijmegen group, 
from the nucleon-nucleon scattering data, does not seem to be 
compatible with their own PWA. Part of the answer comes from the fact 
that in Refs.~\cite{RH04,EM02} the expressions for the $O(q^4)$ NN 
potential were used, while in the Nijmegen work, the $O(q^3)$ NN 
potential was employed. This fact is illustrated in Fig.~\ref{fig:fig5}, 
where in the dashed, light curve the Nijmegen values for the LECs were 
used on the $O(q^3)$ expressions for the NN potential. Another possible 
contribution to this discrepancy could be the fact that we are solving 
the usual Schr\"odinger equation, while the Nijmegen group employs the 
relativistic Schr\"odinger equation in their analysis. This may account 
for the remaining discrepancy and is a topic yet to be investigated. 

\vspace{-3mm}
%\vspace{-3mm}

\section{Acknowledgements}

R. H. would like to thank the organizers of the 18th International IUPAP 
Conference on Few-Body Problems in Physics for the excellent conference 
and opportunity to present this talk.

\end{document}